\documentclass[modern]{aastex62}
\usepackage{rotating}
\usepackage{lineno}
\usepackage{makecell}
\usepackage{soul}
\usepackage{xcolor}
\usepackage[graphicx]{realboxes}
\usepackage{subfigure}
\usepackage{array,multirow}
\usepackage[T1]{fontenc}
\usepackage{amsmath}
\usepackage{makecell}
\usepackage[toc,page]{appendix}

\newcommand{\rpm}{\raisebox{.2ex}{$\scriptstyle\pm$}}

\graphicspath{{./}{figures/}}

\begin{document}

\title{The visual lightcurve of comet C/1995 O1 (Hale-Bopp) from 1995 - 1999}

\author[0000-0003-4659-8653]{M. Womack}
\affil{Florida Space Institute, University of Central Florida, Orlando FL USA}
\affil{Department of Physics, University of Central Florida}

\author[0000-0002-0212-4563]{O. Curtis}
\affil{Institute for Astrophysical Research, Boston University, Boston, MA 02215, USA}

\author{D.A. Rabson}
\affil{Department of Physics, University of South Florida, Tampa, FL 33620, USA}
\affil{National Science Foundation, Alexandria, VA 22314, USA}
\author{O. Harrington Pinto}
\affil{Department of Physics, University of Central Florida}
\author[0000-0002-4884-9367]{K. Wierzchos}
\affil{Catalina Sky Survey, Lunar and Planetary Lab, University of Arizona, Tucson, AZ 85721}
\author[0000-0001-8546-7459]{S. Cruz Gonzalez}
\affil{Department of Physics, University of South Florida, Tampa, FL 33620, USA}
\author[0000-0001-5678-5044]{G. Sarid}
\affil{SETI Institute, Mountain View, CA}
\author[0000-0002-8023-2834]{C. Mentzer} 
\affil{Department of Physics and Astronomy, University of Missouri, Columbia, MO}

\author{N. Lastra}
\affil{Department of Physics and Astronomy, Bowling Green State University, OH}

\author[0000-0003-4236-2053]{N. Pichette}
\affil{Department of Physics, Montana State University, Bozeman, MT}
\author{N. Ruffini}
\affil{Monash University, Australia}
\author{T. Cox}
\affil{Department of Physics, University of South Florida, Tampa, FL 33620, USA}

\author{I. Rivera}
\affil{Department of Physics, University of Central Florida}

\author{A. Micciche}
\author{C. Jackson}
\affil{Department of Physics, University of South Florida, Tampa, FL 33620, USA}
\author{A. Homich}
\author{A. Tollison}

\author{S. Reed}
\author{J. Zilka}
\author{B. Henning}
\author{M. Spinar}
\affil{St. Cloud State University, St. Cloud, MN 56301, USA}

\author{S. Rosslyn Escoto}
\affil{Instituto de Astronomia, Universidad Nactional Autonoma de Mexico, Mexico}
\author{T. Erdahl}
\affil{St. Cloud State University, St. Cloud, MN 56301, USA}
\author[0000-0003-1466-9808]{Marcel P. Goldschen-Ohm}
\affil{Department of Neuroscience, University of Texas at Austin, Austin TX}

\author{W.T. Uhl}
\affil{Department of Astronomy, Yale University, New Haven, CT 06511}
\affil{Dialectic Capital Management, Rowayton, CT 06853}

\correspondingauthor{M. Womack}
\email{mariawomack@gmail.com}, USA

\begin{abstract}

\noindent The long-term brightness evolution of the great comet C/1995 O1 (Hale-Bopp) presented a remarkable opportunity to study the behavior of its coma over four years. We used approximately 2200 total visual magnitudes published in the \textit{International Comet Quarterly}  taken from 17 observers during the period of 1995 July  - 1999 September to create a secular lightcurve. In order to account for observer differences, we present a novel algorithm to reduce scatter and increase precision in a lightcurve compiled from many sources. It is implemented in a publicly available code, ICQSPLITTER. This code addresses the differences among observers by using a self-consistent statistical approach, leading to a sharper lightcurve, and improving the precision of the measured slopes. To first order, the comet's lightcurve approximately follows a $r^{-4}$ response for both pre- and post-perihelion distances. Interestingly, the pre-perihelion data are better fit with a fifth-order polynomial with inflection points at 4.0, 2.6, 2.1 and 1.1 au. We analyze these specific regions and find that they are associated with physical phenomena in the comet's evolution. The first change in slope at 4 au coincides with the development of multiple jets in the inner coma, and the second change in slope at 2.6 au may coincide with the onset of vigorous sublimation of water ice from the nucleus. Outbursts may have occurred a few days before perihelion and at $\sim$ 2.2 and 7.4 au post-perihelion. Contrary to other reports, the lightcurve shows no evidence for the comet having been in outburst at discovery. $Af\rho$ values derived from the visual lightcurve data are consistent with a $r^{-1.5}$ dependence on heliocentric distance, which is similar in shape to those derived from spectroscopy and narrow-band photometry. We present correlation equations for visual magnitudes and CO and H$_2$O production rates, which are consistent with the pre-perihelion visual magnitudes increasing almost entirely due to CO outgassing until a heliocentric distance of about 2.6 - 3.0 au. This is where water production significantly increased. We also present two correlation equations that should prove highly useful for observation planning and data analysis, and can be generalized to be applicable to other comets. 

\end{abstract}


\section{Introduction} \label{sec:intro}

\begin{figure}[ht]
\centering
\includegraphics[width=.3\textwidth, height=0.18\textheight]{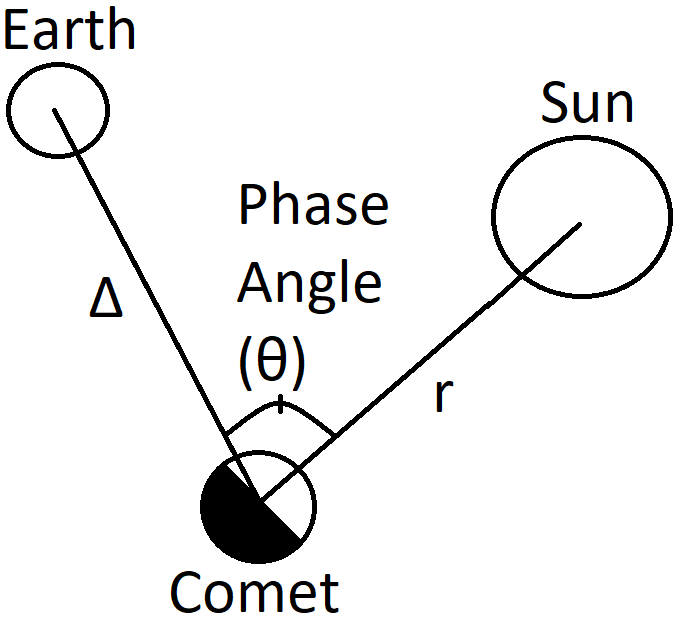}
\caption{A comet's geocentric distance, $\Delta$, and phase angle, $\theta,$ can both affect its apparent brightness.}
\label{fig:phase_angle}
\end{figure}

\noindent Comet nuclei are among the best-preserved icy-rocky remnants from the solar system's formation. Far from the Sun, when comets are inactive, the heliocentric dependence of their bare nuclei's brightness will follow an $r^{-2}$ response due to the reflection of sunlight off the surface, assuming a constant phase angle, $\theta$, surface area and albedo (Fig. \ref{fig:phase_angle}). 

When comets get close enough to the Sun, the ices incorporated in the nucleus sublimate and generate comae of gas and dust. Once they become active, the heliocentric dependence of their brightness changes due to a variety of parameters about which we know very little, including dust particle size distribution, relative contributions of gas and dust, and possibly other sources of energy \citep{fulle2004}. Coma brightness may change when the mass loss rate changes, or large particles or grains in the coma fragment into smaller particles without changing the actual mass loss rate. Moreover, the nucleus's spin pole orientation will affect a comet's heating from the Sun  \citep{whi80}.

Over human history, most reports of comets are from visual observations of their comae, including some reports of comet 1P/Halley $\sim$ 800 hundred years ago, e.g., \citet{choi2018daylight}. Estimates of a comet's visual magnitude are often used to assess its level of activity. For comets with high amounts of dust, lightcurves can also provide useful
information about the bulk behavior of the dust coma, which when coupled with more rigorously
derived values of the dust, and images from infrared, will provide observational constraints for nucleus composition models. Changes in slope may pinpoint a significant transition in the comet nucleus's processing, such as the initiation of distant activity, ramp up of water-ice sublimation, breakup event or outbursts \citep{MeechSvoren04}. 

One of the most continuous datasets on any comet was the visual secular lightcurve of Halley's coma during its 1986 perihelion compiled from thousands of apparent magnitudes and corrected for geocentric distance of the comet ~\citep{green87}. The lightcurve showed many significant changes in slope, peaked at $\sim$ 60 days past perihelion, and continued brighter post-perihelion.

A similar opportunity arose with the ``great'' comet C/1995 O1 (Hale-Bopp), because it was exceptionally bright upon its discovery at $\sim$ 7 au and easily observed for $\sim$ 18 months with the naked eye, and much longer with binoculars and small telescopes. From 1995 - 1999, dozens of observers dedicated nearly every clear night to recording the visual magnitude of Hale-Bopp and submitted brightness estimates to the \textit{International Comet Quarterly (ICQ)}\footnote{\url{http://www.icq.eps.harvard.edu/}} to make them available for further analysis by others.

From the $\sim$ 15,000 magnitudes of Hale-Bopp archived with the ICQ, we studied a subset of 2,240 magnitudes from 17 highly experienced observers. We then created a final lightcurve using a reduced number of  785 magnitudes from 13 observers. In this paper we describe our procedures for removing points, correcting for geocentric distance and phase angle, and our new self-consistent statistical approach for combining different observers, and its efficacy. This reduction process increases the accuracy and precision of a comet lightcurve. 

We present this algorithm through the analysis of the comet Hale-Bopp, which was the subject of the largest observational campaign in history. We also highlight some of the major features and implications of the refined lightcurve. We  present our open source software package \textit{ICQSplitter} \citep{ICQSplitter}, which was used to develop these results and is discussed in Appendix \ref{sec:intro}. This serves as the companion paper to our archive submission with the NASA Planetary Data Systems Small Bodies Node archive \citep{womack18}. 

\bigskip

\section{Observations and Reductions} \label{sec:obs}

\subsection{Pruning data points from the original dataset}\label{sec:removed}

\noindent The original apparent visual magnitudes, $m_{\rm app}$, used in this study were from the ICQ issues 96-100 and provided by Daniel W. E. Green (personal communication), who selected observers with good coverage for both hemispheres (see Table \ref{tab:obs}). Initially, all 2240 of the visual magnitudes were used. We removed 751 points following the best practice guidelines from \cite{green87} and Green 2010 (personal communication), which included removing values when any of the following occurred:

\begin{itemize}
 \item No magnitude method was listed
 \item Methods of obtaining magnitudes that were not the Vsekhsvyatskij-Stevenson-Sidgwick (S), Van Biesbroeck-Bobrovnikoff-Meisel (B), Modified Out-Out, In-focus, or Extrafocal-Extinction methods
 \item Poor observing conditions
 \item Under 20 degrees elevation with no extinction correction applied
 \item Telescopes used for $m_{\rm app} = 5.5$ and brighter
 \item Binoculars used for $m_{\rm app} = 1.5$ and brighter
 \item CCD or photoelectric detectors used
 \item Poor quality star comparison catalogs used 
 \item More than one magnitude per day submitted by an observer (if this occurred, we kept the measurement made with the smallest aperture instrument, which is believed to be most accurate)
\end{itemize}

Some observers, such as Herman Mikuz, submitted a mixture of visual and CCD magnitude estimates. When used properly, CCD and photoelectric detectors can provide very accurate values.  However, they do not always agree well with visual magnitudes determined by eye, and it is not straightforward to combine data obtained with the two techniques (see \citet{kidger02, SosaFernandez2009, almeida2009}). Thus, we removed all CCD measurements from this dataset. Since CCDs were not widely used during the passage of Hale-Bopp, this affected relatively few points.

The remaining apparent magnitudes are the starting point for our analysis and are shown in Fig. \ref{fig:mainlightcurve}a, which plots magnitude vs. heliocentric distance.  The other panels $b, c$ and $d$ of these figures 
show the lightcurve after correcting for geocentric distance, $\Delta$, phase angle, $\theta$, and observer bias, respectively, as we describe in the next sections. To facilitate comparison with other published studies the data are also shown in slightly different format in Fig.  \ref{fig:days_from_peri}a, which plots magnitudes as a function of ``days from perihelion.''

\subsection{Orbital Parameters that Affect the Apparent Brightness of a Comet} \label{sec:params}

The changing orbital configuration of Hale-Bopp, Earth and the Sun sometimes caused the comet's brightness to change over time, which if uncorrected, would be mistakenly attributed to an intrinsic change in output. A comet's apparent visual luminosity, $L_{\rm app}$, is often written in terms of its absolute luminosity at 1 au from the Sun and Earth, $L_0$, scaled with a power-law based on its distance to the Earth ($\Delta$ in au) and Sun ($r$ in au),
 
\begin{equation}
    \label{eq:brightness}
    L_{\rm app}=\frac{L_0} {r^n\Delta^k}
\end{equation}

\noindent where $k$ describes the geocentric distance dependence, which is usually assumed to be ``2'' due to reflected sunlight, and $n$, which is called the ``activity index,'' and is meant to quantify the comet's behavior with respect to heliocentric distance. For a relatively new comet, $n$ = 4 is often used, but the value may change during a comet's orbit and can vary between 1 - 8 \citep{meiselmorris76, MeechSvoren04}. In this paper we assume $k$ = 2, and fit models to the lightcurve data to obtain $n$. 

Using the luminosity-magnitude equivalency equation:
\begin{equation}
    \label{eq:magnitude}
    m_{\rm app}-m_0=2.5\log_{10}(L_0/L_{\rm app}),
\end{equation}
    
\noindent and assuming $k$ = 2, we can rewrite Eq.  \ref{eq:brightness} in terms of magnitudes

\begin{equation}
    \label{eq:magconventional}
    m_{\rm app}=m_0+5\log_{10}\Delta+2.5n\log_{10}r,
\end{equation}

\noindent where  $m_0$ is referred to as the absolute magnitude, or how bright a comet would appear if at 1 au from the Earth and Sun for the case where the comet's phase angle, $\theta$, is ignored (Fig. \ref{fig:phase_angle}).

The visible light from the dust coma includes both reflection and scattering of sunlight off of dust particles in the coma, as well as a much smaller contribution from the comet nucleus. Light scattering off dust particles is a complicated process that depends on the physical properties of the particles in the coma, and is wavelength-dependent  \citep{kolo04}. When observers do correct visual magnitudes for  scattering, they often assume a linear response, which is appropriate for comets with small phase angles (less than $\sim$ 30 degrees) \citep{meechjewitt87, meech04, kolo04}.  

A further complication of correcting for phase angle scattering of visual observations is that many comae observed at visual wavelengths have significant contributions from C$_2$, CO$^+$ and CN gaseous emission, and this light cannot be separated from contributions from the dust without using spectroscopy or narrow-band filters. Hale-Bopp was a very dust-dominated comet, and thus we assumed that all of the light reported by the observers was from dust reflection and scattering. We used the phase correction fuction, $\phi[\theta]$, normalized to 0$^{\circ}$, from \citep{schleicher11}\footnote{Tabulated values of the composite phase function are at this website: \url{http://asteroid.lowell.edu/comet/dustphase.html} and stored at NASA PDS \citep{womack18}.}, which is a composite of a previous fit for comet Halley photometry \citep{Schleicher98}, for smaller angles, and the phase function derived in \cite{marcus07a,marcus07b} for mid- and large-phase angles, which is appropriate for the large heliocentric range of the Hale-Bopp lightcurve. Including the phase angle correction term leads to

\begin{equation}
    \label{eq:phase}
    m_{\rm app}=m_0+5\log_{10}\Delta+2.5n\log_{10}r+2.5\log_{10}(\phi[\theta]).
\end{equation}

Magnitudes corrected to a geocentric distance, $\Delta$, of 1 au are referred to as $m_{\rm helio}$ and are calculated with 

\begin{equation}
\label{eq:geocentric_correction}
m_{\rm helio} = m_{\rm app}-5\log_{10}\Delta
\end{equation}

\noindent using the geocentric distance of the comet at the time the observation is made. Magnitudes are further corrected to a phase angle, $\theta$, of 0 degrees, which we call $m_{\rm phase}$, using

\begin{equation}
\label{eq:phase_correction}
m_{\rm phase} = m_{\rm helio} - 2.5\log(\phi[\theta]).
\end{equation}

\begin{figure}
\centering
\includegraphics[width=0.95\textwidth, height=0.90\textheight]{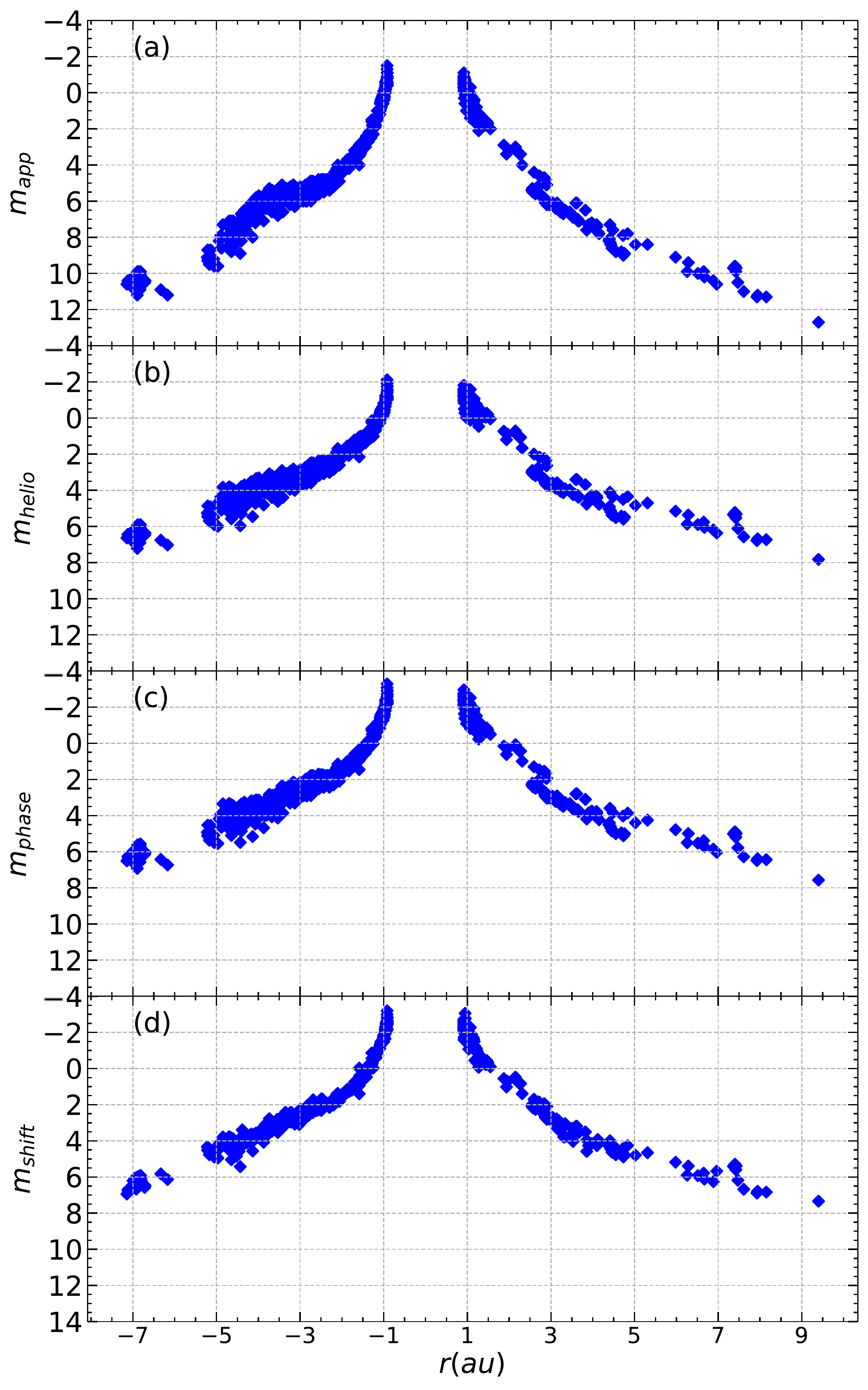}
\caption{Comet Hale-Bopp's lightcurve plotted against perihelion distance. (a) Step 1: $m_{\rm app}$, the original apparent magnitudes after pruning; (b) Step 2: $m_{\rm helio}$, after correcting $m_{\rm app}$ for geocentric distance; (c) Step 3: $m_{\rm phase}$, after correcting $m_{\rm helio}$ for phase angle, and (d) Step 4: $m_{\rm shift}$, the final values after applying the statistical shifts to the $m_{\rm phase}$ values.}
\label{fig:mainlightcurve}
\end{figure}

\begin{figure}
\centering
\includegraphics[width=0.95\textwidth, height=0.9\textheight]{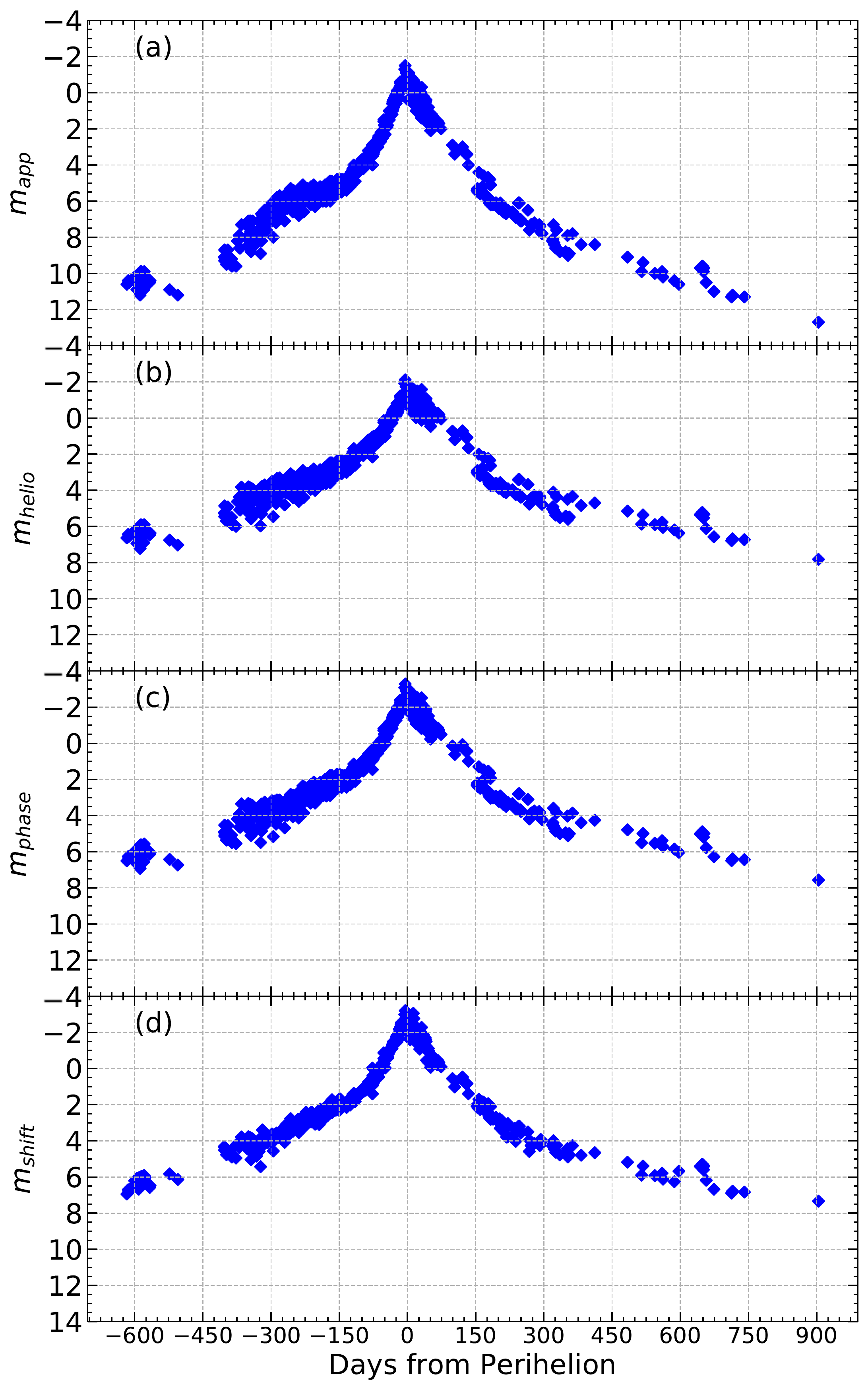}
\caption{Comet Hale-Bopp's lightcurve at every step plotted against days from perihelion. This has the same structure as Fig. \ref{fig:mainlightcurve}.}
\label{fig:days_from_peri}
\end{figure}

We plot the lightcurve for each step in individual panels of Figs. \ref{fig:mainlightcurve} and \ref{fig:days_from_peri}.

\begin{figure}
\centering
\includegraphics[width=0.95\textwidth, height=0.7\textheight]{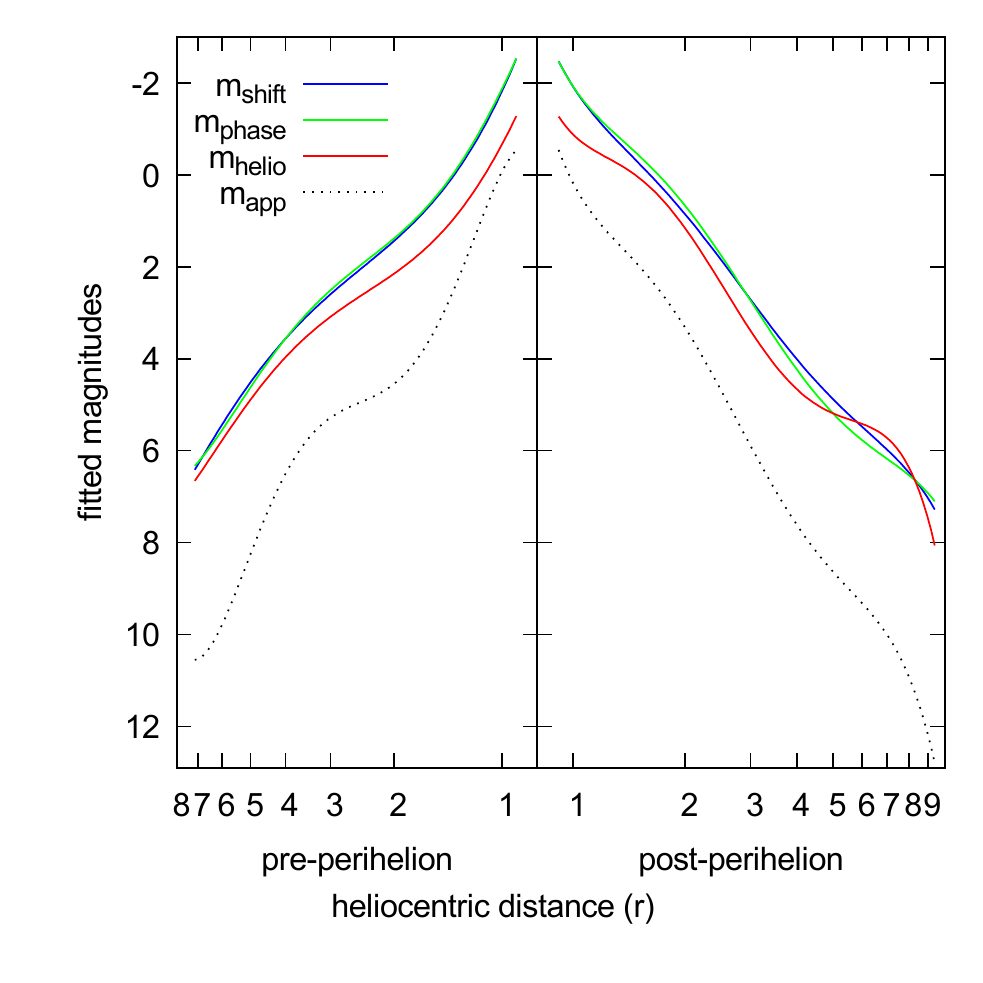}
\caption{Fifth-order polynomial fits to the Hale-Bopp lightcurve data for each step of data reduction. Correcting for geocentric distance (red) and phase angle (green)  makes noticeable changes to the lightcurve shape and generally increases the comet's brightness.}
\label{fig:lcfits}
\end{figure}

\bigskip

\subsection{Statistical analysis of differences among the observers} \label{sec:stats}

In analyzing visual data from multiple observers, the questions inevitably arise
of which data to reject, and under what justification, and whether combining data
from observers, each with his or her own systematic errors, leads to a biased result.
Without instrumental calibration, there is no certain answer to these questions, but as
discussed by \citep{kidger02, mousis14}, such calibration is itself problematic,
and in any case is not available for the observations discussed here.

We offer a systematic approach to combining data from multiple observers yielding a self-consistent consensus fit. In application to comet Hale-Bopp, the
procedure does not significantly affect the grossest measure, the activity index $n$, but, applied to data already corrected for geocentric distance and for phase, does reduce the statistical error bars.

We assume three categories of errors:
\begin{enumerate}
\item Every observer reports the brightness of an object on a scale that is shifted up
or down from other observers, but by the same number of magnitudes, $\delta_{\rm obs}$,
independent of distance or brightness. Without instrumental calibration, we
can best estimate $\delta_{\rm obs}$ as that observer's mean deviation from a consensus fit to
the data (that is, an average).
\item Some observers may have a slope bias, underestimating the brightness of dimmer
objects and overestimating those of brighter ones, or vice-versa. While it is
difficult to correct for such error without calibration, the bias can be detected
(relative to the consensus fit), and that observer's data discarded.
\item Finally, some observers may have a great deal of scatter in their data but no
bias. We can weight these observations less in the fits.
\end{enumerate}

\begin{table} 
\centering

\caption{List of observers whose ICQ data were analyzed. We also include the final shift, $-\delta_{\rm obs}$,  applied to their magnitudes, based on the statistical analysis discussed in section \ref{sec:stats}. Cells marked ``n/a'' signify that the observer too few observations for statistical analysis or that the observations were determined to bias the slope of the lightcurve (against log distance) relative to the consensus and so were discarded in the final iteration.}
\label{tab:obs}
\begin{tabular}{||l l r r l||} 
 \hline
 Observer Name & ICQ Code & shift
 & shift &\\[0.5ex]
& &  pre-perihelion & post-perihelion &\\ [0.5ex] 
 \hline\hline
 Nicolas Biver & BIV & -0.281 & -0.230 &\\ 
 \hline
 John Bortle & BOR & n/a & -0.191 &\\
 \hline
 Reinder Bouma & BOU & 0.145 & n/a & \\
 \hline
 Kazimieras Cernis & CHE03 & n/a & 0.337 & \\
 \hline
 Daniel W.E. Green & GRE & n/a & n/a &\\ 
 \hline
 Werner Hasubick & HAS02& n/a & n/a &\\
 \hline
 Kamil Hornoch & HOR02& n/a & n/a & \\
 \hline
 Albert Jones & JON& -0.597 & n/a &\\
 \hline
 Gary Kronk & KRO02& n/a & 0.165 & \\
 \hline
 Herman Mikuz & MIK& -0.479 & n/a & \\
 \hline
 Andrew Pearce & PEA& -0.029 & n/a &\\
 \hline
 Alfredo Pereira & PER01& 0.084 & 0.241 &\\
 \hline
 Martin Plsek & PLS& n/a & n/a &\\
 \hline
 David Seargent & SEA& 0.439 & 0.399 &\\
 \hline
 Jonathan Shanklin & SHA02& -0.059 & -0.359 &\\
 \hline
 Christopher Spratt & SPR& 0.153 & 0.026 &\\
 \hline
 Vladimir Znojil & ZNO& 0.227 & n/a &\\
 \hline
\end{tabular}
\end{table}

\begin{figure}[ht]
\centering

\includegraphics[width=1.\textwidth, height=0.5\textheight]{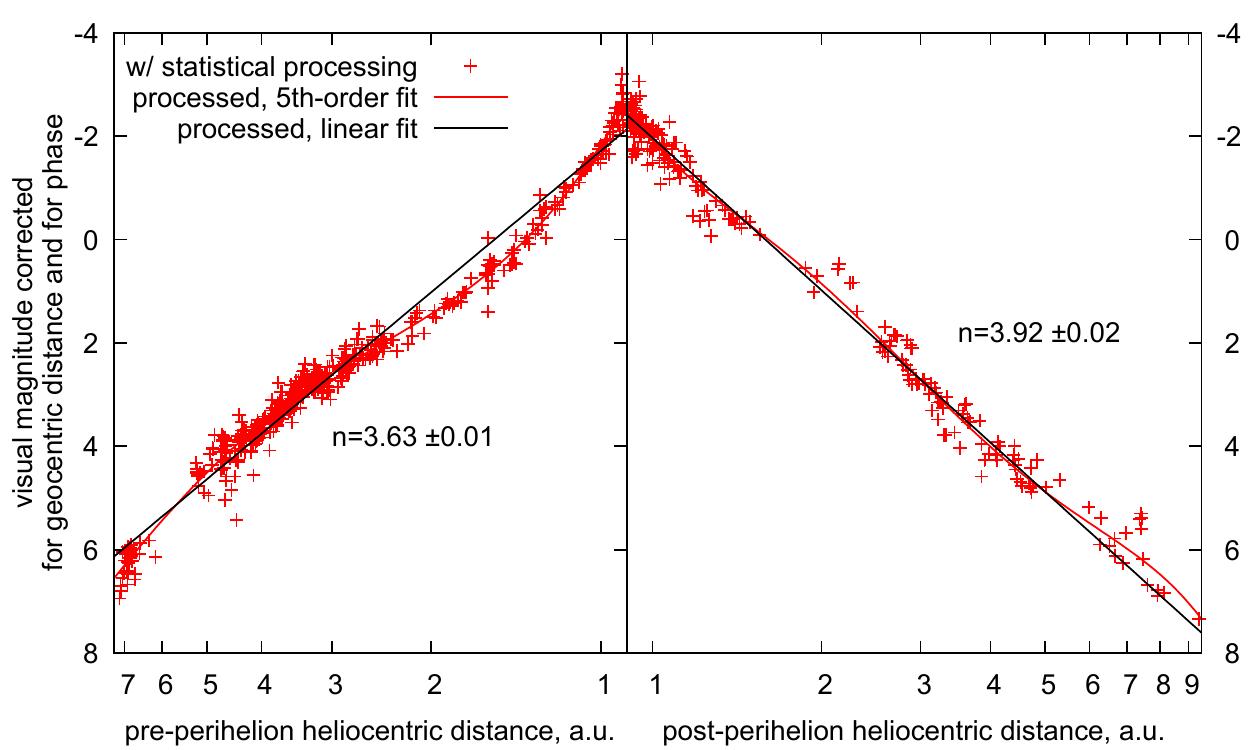}
\caption{Final Hale-Bopp lightcurve: resulting $m_{\rm shift}$ values after applying the statistical process to $m_{\rm phase}$ against a log-$r$ scale. Fits on the graph are processed linear (black line) and 5th-order polynomial fits (red line). 
}
\label{fig:final_logscale}
\end{figure}

We seek a consensus fit to the data.  The comet exhibits small but noticeable deviations from a power law (luminosity $\sim r^{-n}$) on time (distance) scales larger than
any outbursts. In particular, the pre-perihelion data between 2 and 1 au
suggest a positive curvature in the graph of (minus) magnitude against log distance (See Fig.~\ref{fig:final_logscale}.)  A straight-line fit would penalize observations that report this feature accurately.
Based on the total number of apparent features between the closest (0.91 au) and
furthest (9.4 au post-perihelion) observation, we fit to fifth-order polynomials. The
effect is to reflect the smoothed consensus.

The deviations between an observer's measurements and the consensus fit at the
same distance are considered noise. The set of all such deviations by one observer
defines the noise distribution for that observer, characterized by a mean $\delta_{\rm obs}$, variance
$\sigma^2_{\rm obs}$, skewness, excess kurtosis, etc.

To compute the self-consistent fit, we iterate the following until convergence
to a fractional tolerance of 0.0001 (absolute tolerance if any fitting parameter is less
than 0.0001) of all six polynomial fitting parameters:
\begingroup
\renewcommand{\theenumi}{\roman{enumi}} 
\begin{enumerate}
\item We fit a fifth-order polynomial through all the data of the (possibly shifted)
magnitudes against log distance by the method of least squares \citep{legendre1805nouvelles}, weighting each observation inversely as the observer's variance of deviations, $\sigma^2_{\rm obs}$. Initially, the magnitudes have been corrected for geocentric distance and for phase angle, but have not been shifted. Also initially, the weights are 
all equal, because we do not know the distributions of deviations.\label{task:first}
\item  Using the same data as in \ref{task:first}, we can also fit a straight line, recovering the activity index ${n}$ and statistical error bars on that fit based on the shifted, weighted data.
\item For each observer, we consider the distribution of deviations between the observations and the fit at the given distance (i.e., in the graph, the vertical vectors between data points and the polynomial fit). For each observer separately, we treat the distribution as noise and estimate the mean, $\delta_{\rm obs}$, and variance, $\sigma^2_{\rm obs}$.\label{task:dev}
\item For each observer separately, we shift all magnitudes by $-\delta_{\rm obs}$ as computed in \ref{task:dev}.\label{task:last}

\end{enumerate}
\endgroup
For the present data set, these iterations converge to the specified tolerance after between ten and twelve iterations. We now have a self-consistent fit and set of shifted data. Row 4a of Table \ref{tab:stats} shows small changes in best-fit slopes after the procedure, with the statistical error of the straight-line fit cut roughly in half. Note that non-straight-line features in the consensus polynomial fit limit how far the statistical error bars in a straight-line fit can shrink.

\begin{table}[ht]
\centering
\caption{Best-fit activity indices $n$ for pre-perihelion and post-perihelion observations before and after corrections and the self-consistent method.  (The slope of magnitude against $\log_{10}(r)$ is equal to $2.5n$; see Eq. \eqref{eq:magconventional}.)  Steps 4a and 4b use self consistently shifted magnitudes and weight each observer's data inversely as $\sigma_{\textrm{obs}}^2$, the variance of the noise. Error bars are standard errors of fit \citep{NumericalRecipes3} and would be considerably smaller, pre-perihelion, if slow variations (Fig.~\ref{fig:final_logscale}) were subtracted. The fitting method reduces statistical error but does not significantly alter the activity index estimates.
\label{tab:stats}} 
\begin{tabular}{l r r r} 
 & pre-perihelion & post-perihelion &\\
Correction Step & -7.0 to -0.9 au & 0.9 to 8.0 au & \\
 \hline
 1. raw data ($m_{\rm app}$) & 4.53 \rpm 0.03 & 5.07 \rpm 0.03&\\ 
 2. geocentric distance correction only ($m_{\rm helio}$)& 3.22 \rpm 0.02 & 3.66 \rpm 0.03 &\\
 3. geocentric distance and phase correction ($m_{\rm phase}$)& 3.66 \rpm 0.02 & 3.97 \rpm 0.03 &\\
 &  &  &\\
 4a. self-consistent shifts & 3.66 \rpm 0.01  & 3.94 \rpm 0.01 &\\
 4b. drop observers, self-consistent shifts ($m_{\rm shift}$)& 3.63 \rpm 0.01 & 3.92 \rpm {0.02} &\\
 \hline
\end{tabular}
\end{table}

The self-consistent procedure has eliminated the need to discard data arbitrarily (e.g., points more than some number of standard deviations above or below a consensus fit, which otherwise would throw off least-squares fits) by weighting points inversely as the observer's variance. However, as noted above, an observer whose
systematic error changes with brightness would still affect the activity index adversely. We can detect such a systematic non-stationarity by applying Student's $t$-test to each observer's data set, comparing the mean deviation from the consensus fit in
the first half of the observer's data (sorted by distance) to the mean deviation in the
second half. Since the variances of the two halves may not be equal, we normalize
by the ``pooled variance'' to get an approximate $t$-statistic \citep{NumericalRecipes3}. Assuming approximately
Gaussian noise, we calculate the $p$ (probability) value that the $t$-statistic would be as large as observed or larger under the null hypothesis that first and second halves of the data were drawn from the same distribution, i.e., that the observer did not contribute bias to the slope relative to the consensus (stationarity).

For a $p$ value less than 0.05, we reject the null hypothesis and 
conclude, on the basis of the $t$-test, that the observer's data bias the slope relative to the consensus.  We then discard such
observers from the data set and repeat the self-consistent iteration (\ref{task:first}--\ref{task:last}), starting from
the original (phase-corrected, unshifted) magnitudes but without the discarded observations. If
repeating the $t$-test at the end of second set of iterations results in more observers
biasing the slope (relative to the consensus) based on the $t$-test, we then repeat the procedure until no new observers are discarded.
For the pre-perihelion data, one additional observer failed the $t$-test after the second set of
iterations, so we iterated a third time without that observer's data. At the end of the
third set of iterations, no new observers failed the $t$-test.

As shown by Line 4b of Table \ref{tab:stats}, throwing out observers results in small changes
in the slopes. For the pre-perihelion data, seven observers out of 17 were discarded,
accounting for 430 of the original 1,003 data points. For post-perihelion data, four
out of 12 observers were discarded, accounting for 232 out of the original 486 data
points.

For comet Hale-Bopp, the self-consistent procedure resulted in adjustments to the
slope roughly comparable to the original statistical error bars, while cutting the error bars
approximately in half, as reflected in visibly smaller scatter in the lightcurve.  (See Table~\ref{tab:stats} between lines 3 and 4b and Fig.~\ref{fig:mainlightcurve} between panels (c) and (d).)

These changes are far smaller than those associated with correcting for geocentric distance and phase. The observations in the present work were taken by people using generally reliable star comparison catalogs. Future work may rely on less homogeneous amateur networks, in
which case a self-consistent method for combining and weighting magnitudes, and discarding subsets with systematic slope bias, could prove useful.

The shifts, $-\delta_{\rm obs}$, to the data resulting from this self-consistent procedure are listed in Table \ref{tab:obs} and are equivalent to the last step in the equation
\begin{equation}
    \label{eq:shift}
    m_{\rm app}=m_0+5\log_{10}\Delta+2.5n\log_{10}r+2.5\log_{10}(\phi[\theta]) + \delta_{\rm obs},
\end{equation}

\noindent and we define

\begin{equation}
    \label{eq:withshift1}
    m_{\rm shift} = m_{\rm phase} - \delta_{\rm obs},
\end{equation}

\noindent or as it is written in complete form:

\begin{equation}
    \label{eq:mwithshift2}
    m_{\rm shift} = m_{\rm app}-5\log_{10}\Delta - 2.5\log_{10}(\phi[\theta]) - \delta_{\rm obs} .
\end{equation}

\noindent Using Eq. \ref{eq:mwithshift2}, we can rewrite Eq. \ref{eq:shift}  as:

\begin{equation}
    \label{eq:mshiftslope}
    m_{\rm shift}=m_0+2.5n\log_{10}r, 
\end{equation}

\noindent which we use to measure $n$.

The $m_{\rm shift}$ values represent the final corrected magnitudes of the lightcurve and are shown in Figs. \ref{fig:mainlightcurve}d and \ref{fig:days_from_peri}d, as well as a close-up version in Fig. \ref{fig:final_logscale}. 

Correcting for geocentric distance and phase angle makes noticeable changes to the lightcurve shape and increases the comet's brightness. By design, making self-consistent shifts to the data does not change the lightcurve's shape, but reduces the uncertainty. For ease of comparison, fifth-order polynomial fits to the data at each correction step are shown in Fig. \ref{fig:lcfits}. 

The original apparent magnitudes and those at all steps outlined in this paper underwent full panel review with the NASA Planetary Data Systems Small Bodies Node and are publicly archived \citep{womack18}. All data reduction was performed using the ICQSplitter program, which is  available for download on GitHub.\footnote{\url{https://github.com/curtisa1/ICQSplitter}} This paper describes the functionality of ICQSplitter Version 3.0 as of 28 January 2020. Since ICQSplitter is still evolving, refer to the on-line documentation for up-to-date information.  Appendix \ref{sec:ICQSplitter} discusses ICQSplitter's functionality and its subroutines in more detail.

\section{Discussion}

\subsection{Comparison with other Hale-Bopp lightcurves}
\label{subsec:comparison}

Several other groups produced visual lightcurves of Hale-Bopp that spanned many years; however, they were not always derived the same way. For example, some did not account for the effect of geocentric distance, even though this had up to a 42\% impact on the activity index, $n$, for Hale-Bopp. Fig. \ref{fig:mainlightcurve} shows an illustration of this effect, where a ``shoulder'' feature, seen in panel \ref{fig:mainlightcurve}a at $\sim$ 3.5 -- 4.0 au pre-perihelion, is effectively removed once we account for the geocentric distance in panel \ref{fig:mainlightcurve}b. This type of artifact might invite unwarranted physical interpretation.

Table \ref{tab:HBLCs} lists some of these longterm lightcurve data sets for Hale-Bopp. Our lightcurve is the only one in Table \ref{tab:HBLCs} that corrected for light scattering in the coma (see Sec. \ref{sec:params}). Our calculations show that scattering can be significant for Hale-Bopp's lightcurve and at times contributed up to a 92\% change to the calculated brightness (e.g., compare Figs. \ref{fig:mainlightcurve}b and \ref{fig:mainlightcurve}c and Fig. \ref{fig:lcfits}). This result underscores our motivation in sharing the ICQSplitter code, so that this potentially important correction step may be more accessible to other researchers.

The comet magnitude websites from  the {\it Comet OBServational Database}\footnote{\url{https://cobs.si/}} (COBS) and Seiichi Yoshida\footnote{http://www.aerith.net/comet/catalog/1995O1/1995O1.html} provide data and interactive tools to analyze visual and CCD magnitudes submitted to the International Comet Quarterly (ICQ) and the Minor Planet Center (MPC). Both Yoshida and COBS present only apparent magnitudes ($m_{\rm app}$), but include correction for geocentric distance. However, when fitting slopes to their lightcurves, they do not correct for phase angle variations, or include any adjustments for observer differences 
(Yoshida, S. 2020, personal communication, \citet{Zakrajsek2018,filonenko2001}).

The COBS online tool provides greater flexibility by allowing the user to choose which starting and ending dates to use in a fit, whereas Yoshida provides fits to pre-selected temporal ranges. 
Based on our analysis here, we recommend that caution should be used when interpreting any changes in slopes derived from Yoshida and COBS lightcurves. They present uncorrected apparent magnitudes, which might include observational artifacts, as discussed in the above mentioned example for Hale-Bopp.

\begin{table} 
\centering
\caption{\bf Characteristics of visual comet Hale-Bopp lightcurves \label{tab:HBLCs}} 
\begin{tabular}{||l l l   ||} 
 \hline
 {\bf Study} & {\bf Time span} & {\bf Analysis method} \\ 
 \hline\hline
This paper &  1995-1999  & Corrected for $\Delta$ and $\theta$ (coma), self-consistent  \\
& & statistical analysis of magnitudes \\
COBS$^4$ & 1995-2006 &  Corrected for $\Delta$, least squares fit to all data  \\
Yoshida$^5$ & 1995-2013  & Corrected for $\Delta$, least squares fit to all data  \\
\citet{liller2001ccd} &  1995-2000&   Corrected for $\Delta$, fit determined by eye  \\
\citet{kidger97best} &  1995-1997 &  Corrected for $\Delta$, least squares fit to all data \\
\citet{ferrin2010atlas} &  1993-2009  &  Corrected for $\Delta$, $\theta$ (nucleus), least squares fit
\\
& &  to brightest magnitudes from many observers \\
\hline
\end{tabular}

\end{table}

\cite{liller2001ccd} and \citet{kidger97best} both applied geocentric distance corrections, but did not correct for phase angle. \cite{liller2001ccd} constructed Hale-Bopp lightcurves using visual magnitudes from the IAU circulars and derived an activity index through ``fitting by eye''. There was no assessment of potential observer differences or fitting bias. \citet{kidger97best} used visual observations from {\it The Astronomer Group} and {\it The British Astronomical Association Comet Section}. They applied a least-squares fit to their data, but did not consider any adjustments for observer differences. 

Another well known lightcurve was published by \cite{ferrin2010atlas}, which includes geocentric distance correction and adjustment of visual magnitudes due to scattering. However, scattering was considered for the nucleus only. This constitutes a very small change for Hale-Bopp's nucleus, since it was surrounded by a significant coma.      

Since all of the lightcurves in Table \ref{tab:HBLCs} were corrected for geocentric distance (which we refer to as Step 2), we briefly compare their activity index, $n$, for this Step of the lightcurve calculations, using 

\begin{equation}
    \label{eq:mhelioslope}
    m_{\rm helio}=m_0+2.5n\log_{10}r, 
\end{equation}

\noindent which we derive from Eqs.  \ref{eq:magconventional} and \ref{eq:geocentric_correction}\footnote{Two conventions are used to describe lightcurves and, unfortunately, both use the letter $n$. COBS, \citet{ferrin2010atlas}, \citet{liller2001ccd}, and this paper calculate activity indices, $n$, according to Eq. \ref{eq:magconventional}, whereas Yoshida and \citet{kidger97best} use $n$ for the combined term $2.5n$ in Eq. \ref{eq:mhelioslope}. When we mention results from Yoshida and \citet{kidger97best}, we have already divided their $n$ values by 2.5.}.

Overall, the Step 2 lightcurves for the groups in Table \ref{tab:HBLCs} agree within the uncertainties.
The online tools provided by Yoshida and COBS, and \citet{liller2001ccd} lightcurves show the best agreement with our Step 2 values. This is not surprising considering all four teams started with ICQ measurements. Yoshida and COBS give activity indices that are within $\sim$ 10\% of each other's values, Liller's ``fit by eye'' value, and our  measurements. Our lightcurve is also within 2-6\% agreement with \citet{kidger97best}, except for 2.6-2.1 au (1996 Oct 25 - Dec 4), when their calues are $\sim$ 30\% higher than ours, and during 1.1-0.9 au (1997 Mar 1 - Apr 1) when their $n$ is $\sim$ 60\% of our index. However, the data have high disperion in both of these regions in \citet{kidger97best} and they overlap with our values within their uncertainties. \citet{ferrin2010atlas} present two possible activity indices within 6.3 au of the Sun that are $\sim$ 20-30\% flatter than everyone else's value for this long range. Their lower index may be due to the fact that they only fit to the 5\% brightest points and combine visual and CCD-derived magnitudes. \citet{ferrin2010atlas} incorporate naked-eye visual magnitudes from the ICQ as well as CCD measurements, so many individual observers contributed results. As we have shown in this paper, some observers submitting to the ICQ are not self-consistent, and some submit systematically brighter, or lower, magnitudes than the average. Thus, it may be that preferentially fitting the brightest points when multiple observers are involved contributed to biases in \citet{ferrin2010atlas}'s results for Hale-Bopp.

\subsection{Pre-perihelion Changes in the Lightcurve Slope}
\label{subsec:changes}

Given the significant effect that the phase angle correction can have on coma brightness, and the benefits of using a self-consistent method for adding data from different observers, the rest of our analysis is completed using our final lightcurve data (Step 4).

Comet Hale-Bopp's lightcurve provides information about the long-term response of the nucleus to solar heating and continuous thermal processing at different levels. Additional information about the nucleus, such as shape, rotation, or active regions is needed to constrain models in detail \citep{Marshall2019water}. However, long-term lightcurves with higher resolution and accuracy are useful in constraining the parameter space for nucleus thermal models, and provide an important way of connecting models of coma activity and nucleus processing, especially over multiple orbits ot larger arcs of a given orbit. A change in the lightcurve's shape may indicate the onset of sublimation or outgassing of a new volatile, or pinpoint the timing of prominent outbursts. 

To first order, the activity index of the entire pre-perihelion lightcurve is $n$ = 3.63 $\pm$ 0.01 (Fig. \ref{fig:final_logscale}). This is within $\sim$10\% of the $r^{-4}$ value that is often assumed for relatively new, inbound comets \citep{meech04}. It is also consistent with nucleus models that are under the influence of a CO-dominated outgassing regime with a possible water-ice phase change from amorphous to crystalline state \citep{prialnik97}. 

As Fig. \ref{fig:final_logscale} shows, however, the lightcurve noticeably deviates from a straight line in several places, and a fifth order polynomial is a better fit with inflection points at $r=$ 4.0, 2.6, 2.1 and 1.1 au. The locations of these changes in slope were also noted by \citet{kidger97best}. In Table \ref{tab:slopes} we list the power-law activity index for the five regions at each of the data reduction steps.  Next, we discuss the lightcurve changes in the context of other coma phenomena during the same time.

\begin{table}[h!]    
\centering
\caption{Hale-Bopp's pre-perihelion lightcurve showed changes in slope at $r$ = 4.0, 2.6, 2.1, and 1.1 au. This table lists the activity index ($n$) and absolute magnitude ($m_0$) fits for each region. Corrections for geocentric distance  and phase angle can produce significant changes in slopes. 
The preferred activity index for each heliocentric range are listed in bold font. }

\label{tab:slopes}
\begin{tabular}{||c l l c  c ||} \hline
Heliocentric  & Dust coma description & Step & n &  $m_0$   \\ 
region (au) & &  & &\\ \hline
\multirow{4}{*}{A) 7.2 - 4.0} & &$m_{\rm app}$ & 6.75 \rpm 0.20 &  -3.57 \rpm 0.36 \\
&{Single jet transitions to} & $m_{\rm helio}$ & 4.27 \rpm 0.18  & -2.48 \rpm 0.32   \\
&{multiple jets at 4 au}& $m_{\rm phase}$ & 4.39 \rpm 0.18 &  -3.03 \rpm 0.33  \\
& & $m_{\rm shift}$ & \bf{4.46 \rpm 0.11}  & -3.15 \rpm 0.20  \\ \hline

\multirow{4}{*}{B) 3.9 - 2.6} &  &$m_{\rm app}$ & 1.18 \rpm 0.17 & 4.11 \rpm 0.22  \\
& Multiple jets maintained, &$m_{\rm helio}$ & 1.62 \rpm 0.16  & 1.26 \rpm 0.21  \\
& outburst near 3.1 au&$m_{\rm phase}$ & 3.11 \rpm 0.17 &  -1.19 \rpm 0.22  \\
& &$m_{\rm shift}$ & \bf{3.06 \rpm 0.10} &  -1.10 \rpm 0.13  \\ \hline

\multirow{4}{*}{C) 2.6 - 2.1} & & $m_{\rm app}$ & 3.42 \rpm 0.76 &  1.71 \rpm 0.73 \\
&Multiple jets maintained& $m_{\rm helio}$ & 2.94 \rpm 0.75 &  -0.22 \rpm 0.72  \\
&& $m_{\rm phase}$ & 2.29 \rpm 0.76 &  -0.23 \rpm 0.73  \\
&& $m_{\rm shift}$ & \bf{2.36 \rpm 0.47}  & -0.29 \rpm 0.45  \\ \hline

\multirow{4}{*}{D) 2.0 - 1.1} & Curved jets, gull wing &$m_{\rm app}$ & 4.79 \rpm 0.20 &  0.76 \rpm 0.10 \\
&and ring arc features  & $m_{\rm helio}$ & 2.79 \rpm 0.19  & -0.05 \rpm 0.09  \\
&appear at $\sim$1.2 au& $m_{\rm phase}$ & 3.77 \rpm 0.20 &  -1.26 \rpm 0.09 \\
&& $m_{\rm shift}$ & \bf{4.01 \rpm 0.15} &  -1.41 \rpm 0.07  \\ \hline

\multirow{4}{*}{E) 1.0 - 0.9} && $m_{\rm app}$ & 8.63 \rpm 0.52 &  0.10 \rpm 0.04 \\
&Perihelion surge/outburst& $m_{\rm helio}$ & 7.19 \rpm 0.53  & -0.63 \rpm 0.04  \\
&and ring arc features continue& $m_{\rm phase}$ & 7.50 \rpm 0.53 &  -1.78 \rpm 0.04 \\
&& $m_{\rm shift}$ & \bf{7.52 \rpm 0.40}  & -1.81 \rpm 0.03  \\ \hline

\end{tabular}
\end{table}

\subsubsection{Region A: 7.2 $\ge$ r $\ge$  4.0 au (1995 July 17 -- 1996 June 23)}


Several observers reported  1-2 magnitude outbursts soon after discovery at $\sim$ 7 au \citep[see discussion in][]{prialnik97} and near 4.0 au \citep{liller2001ccd}, but we see no evidence for either in our lightcurve. The activity index for region A is $n$ = 4.46 $\pm$ 0.11. The comet was still far enough from the Sun that water-ice sublimation, the dominant activity mechanism for most comets, was inefficient. The coma was probably controlled by outgassing of CO, CO$_2$, and other volatiles \citep{jew96,wom97, cro99,biv02}, and possibly moderated by crystallization of amorphous water ice  \citep{prialnik97}. 

During this time, optical images  show that Hale-Bopp's dust coma exhibited a single radial jet feature which persisted for several months and then transitioned to four or more jets at $\sim$ 4 au \citep{braunstein97,Mueller1997}. The appearance of the jets coincides with the next change in lightcurve slope and could indicate the onset of several active areas on the nucleus \citep{Sek97}, perhaps related to triggering of subsurface volatile pockets, or mass wasting events due to mechanically or thermally unstable regions on the nucleus.

\subsubsection{Region B: 4.0 > r $\ge$   2.6  au (1996 Jun 23 -- October 25)}

Here, the slope flattens with n = 3.06, which it maintained for about four months. Previous analyses by \cite{kidger97best} reported a much flatter $r^{-1.5}$ brightening law for this region, which was attributed to the onset of water-ice sublimation near 3 au \citep{BockeleeMorvanRickman}. However, our calculations show that the flattening did not have a physical cause, but instead was due to a prominent and uncorrected phase angle effect. This is clearly seen if we compare the indices for the geocentric-corrected ($n$ = 1.62) and phase-corrected ($n$ = 3.11) lightcurves of region B in Table \ref{tab:slopes}. 


An outburst was reported in mid-September (3.1 au), when at least a few of the jets appeared disrupted \citep{braunstein97, Tao2000HBOutburst, mccarthy2007}. This outburst did not appear in our lightcurve, nor did it cause a change in the lightcurve, as the coma maintained a slope of $n\sim$ 3 for another month until $\sim$ 2.6 au (mid-late October 1996). This slope change is within the heliocentric distance where it is is commonly thought that water-ice sublimation becomes prominent. If rapid water-ice sublimation is initiated it can easily surpass outgassing of all other volatile species. Due to increased thermal flux and spatial distribution throughout the nucleus, water-ice quickly becomes the primary carrier of dust grains \cite{prialnik97}, which can significantly increase the reflectance behavior of the coma. 

\subsubsection{Region C. 2.6 >r $\ge$ 2.1 au (1996 October 25 - December 4)}


In this region, the lightcurve flattened even more with a $r^{-2.36}$ trend. During this time, the comet maintained at least 4 nearly-radial jets in the inner coma, possibly from the same active areas that started at the end of region A. If indeed supported by subsurface pockets or newly excavated patches, this sustained jet activity can inform us about the volume of volatile species (mostly water ice) and how accessible it is to sublimation and distinct outward flow.

\subsubsection{Region D. 2.1 > r $\ge$ 1.1 au (1996 December 5 - 1997 February 25)}


At $\sim$ 2 au, the lightcurve returns to a $r^{-4}$ brightening law.  No outbursts are detected. During this time the inner coma morphology transitioned once again. Between 1996 Nov 16 (2.33 au) and 1997 Feb 13 (1.23 au) the coma's dominant features changed from the nearly radial jets to curved jets consistent with radiation pressure and  beginning of a ``gull wing'' feature  \citep{braunstein97}. Modeling and simulations of many optical and infrared images indicated that the changes from jet-like to arc-like features were due to a changing observing perspective and increased angular resolution, and not due to physical changes in the nucleus or coma \citep{samarasinha97}. This again can be considered consistent with a view of water-ice rich activated pockets that are present just below the surface.

\subsubsection{Region E. 1.1 > r $\ge$ 0.9 au (1997 Mar 1 - Apr 1)} 


Here we see the fastest rate of brightening of any region, with a power-law index of $r^{-7.5}.$ The extreme brightening here compared to other regions has been interpreted to mean that an outburst occurred during the dates leading up to perihelion \citep{samarasinha97}. The brightest points in the lightcurve occurred 3 days before perihelion, in agreement with previous analysis \citep{ferrin2010atlas, kidger97best}. During this time the optical images continued to show the inner coma features dominated by curved jets with shells, or arcs. 

\subsection{Post-perihelion lightcurve}

Instead of a series of changing slopes, the post-perihelion data are well-fit by a $r^{-3.94 \pm 0.02}$  power law superimposed with some outbursts. We did not notice the breaks claimed by \cite{kidger97best}, nor do we see any evidence that the comet faded more rapidly as water-ice sublimation turned off around 3 au, which they reported.  The lack of measurable slope change over 8 au is consistent with the comet nucleus achieving a level of stability after its perihelion passage. This is slightly ($\sim$8\%) steeper than the overall change measured over the pre-perihelion lightcurve from 7 to 0.9 au. 

Outbursts may have occurred at 2.15 $\pm$ 0.07 au and 7.36 $\pm$ 0.20 au (Fig. \ref{fig:final_logscale}), both lasting a few weeks. The 2.15 au outburst in 1997 Jul led to a brightness change of 0.47 $\pm$ 0.08 magnitudes. The 7.36 au outburst in 1999 Jan  brightened the coma by 0.68 $\pm$ 0.23 magnitudes, and was also observed by \citet{liller2001ccd}'s CCD dataset, who measured an increase of $\Delta m \sim$  1.1 magnitude.


\subsection{Absolute magnitude}

Absolute magnitudes are a theoretical construct, representing what brightness the comet would have had if viewed at 1 au from both the Sun and Earth. Astronomers typically correct the apparent magnitude to 1 au from the Earth with Eq. \ref{eq:geocentric_correction}, but the correction to 1 au from the Sun is often  model-dependent. However, since Hale-Bopp passed and was observed twice through the heliocentric distance of 1.0 au, we have the opportunity to measure the absolute magnitude directly. This is valuable for constraining models of comet activity. Using all the $m_{\rm shift}$ values that were within 0.05 au of 1.0 au, we calculate these average absolute magnitudes: 
\medskip

Pre-perihelion: $m_0$ = -1.74\rpm{0.13}

Post-perihelion: $m_0$ = -1.92\rpm{0.23}

\medskip

\noindent where the uncertainties are one standard deviation. Within these uncertainties we can see that the comet was equally bright pre- and post-perihelion at 1 au.  These absolute magnitudes confirm that Hale-Bopp was one of the brightest comets ever recorded.

\subsection{$Af\rho$ Calculations}

$Af\rho$ is a quantity often used as a proxy for dust production rate, and it can be thought of as equivalent to the size of a hypothetical disk needed to approximate the light reflected from the dust at the comet's distance. It is derived from $A$, the albedo of the dust, $f$, the filling factor, and $\rho$, the coma radius \citep{ahearn84} 

\begin{equation}
\label{eq:afp}
Af\rho = \frac{(2r\Delta)^2}{\rho}\frac{F_{\rm comet}}{F_{\rm Sun}},
\end{equation}

\noindent where 
\begin{equation}
\label{eq:afp2}
F = 10^{-\frac{m}{2.512}}
\end{equation}

\noindent is used to convert magnitudes, $m$, to flux, $F$, and we assume $m_{Sun} = -26.7$. We used a value of half the observers' reported coma diameter for the aperture radius, $\rho$.

To our knowledge, these are the first reports of $Af\rho$ calculated from visual magnitudes for Hale-Bopp, or any comet. The quantity is generally not considered to be well-suited for visual magnitudes, since an aperture radius is usually not provided by the observers. Also, as discussed earlier, visual magnitudes  recorded without a filter (which apply to all the data in our study) may be contaminated with gaseous emission, especially within $\sim$ 2 au of the Sun, when gas output increases significantly.  Nonetheless, since we could not find evidence of this being done before, we calculated $Af\rho$ values from the final shifted visual magnitudes (with the geocentric correction reversed, since Eq. \ref{eq:afp} accounts for geocentric distance).  

\begin{figure}
\centering
\includegraphics[width=.85\textwidth, height=0.4\textheight]{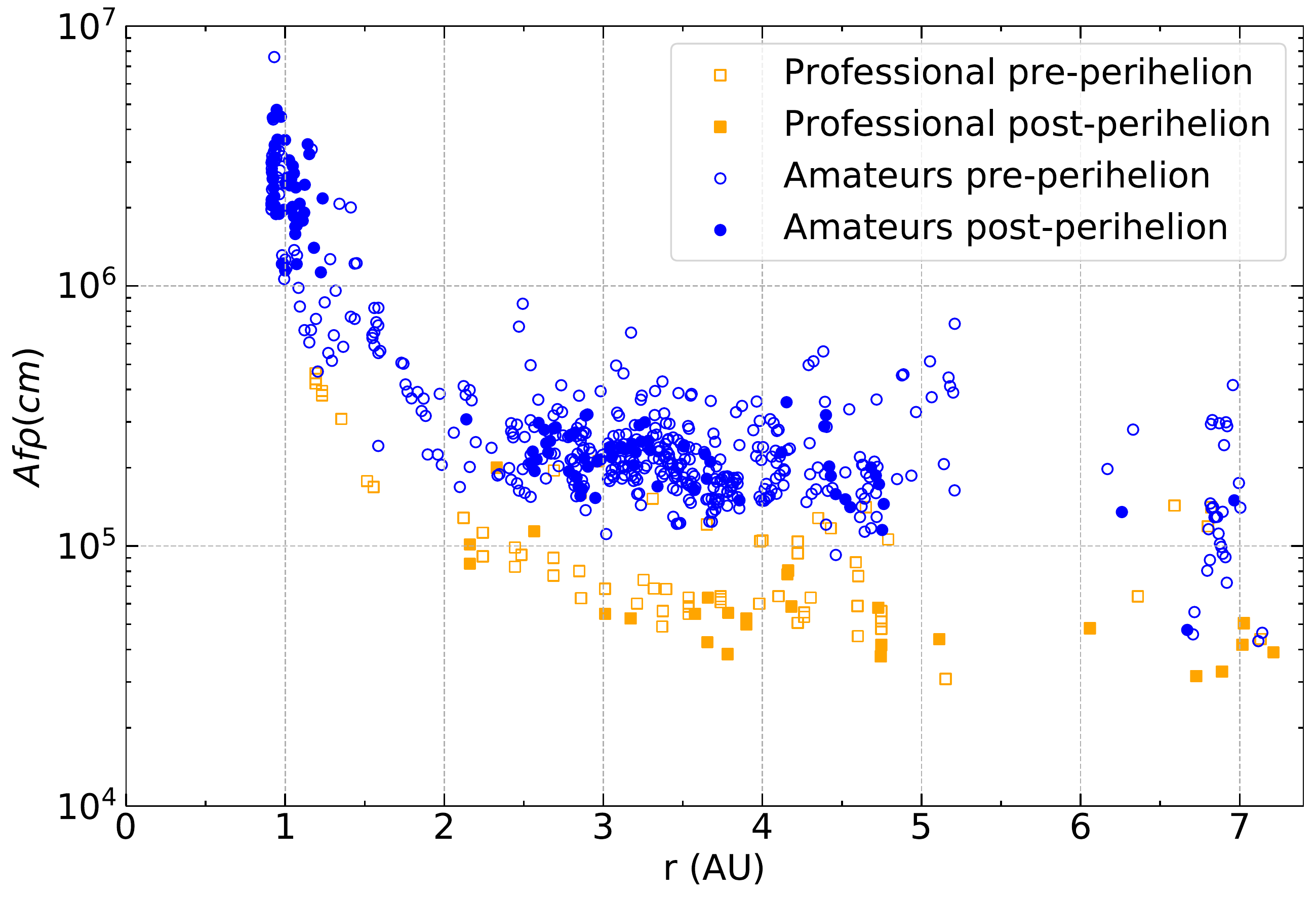}
\caption{$Af\rho$ values (in cm) derived from our final Hale-Bopp lightcurve data (blue circles) and $Af\rho$ values derived from optical spectroscopy and R-filter images (orange squares) from Fig. 4 of \citet{weiler03}. Open symbols are used for pre-perihelion data and filled symbols denote post-perihelion. While there is some overlap, in general, the $Af\rho$ values derived from visual magnitudes are frequently 2-4 times higher, on average, than those derived using optical spectroscopy and R-filter imaging. Both sets show that the $Af\rho$ values have a strong inverse relationship with heliocentric distance within 3 au, but appears to change very little beyond that. 
}
\label{fig:afp}
\end{figure}

In Fig. \ref{fig:afp} we plot our magnitude-determined $Af\rho$ values along with the values derived from spectroscopic and filter-imaging techniques, using data from Fig. 4 from \citet{weiler03}. As Fig. \ref{fig:afp}  shows, the values show a very similar trend in the pre- and post-perihelion data. When combined, our pre-perihelion and post-perihelion values follow a $Af\rho \sim$ 2x10$^6$r$^{-1.5}$ cm heliocentric dependence. No heliocentric dependence was provided by \citet{weiler03}, but both $Af\rho$ curves show broadly similar shapes with a steep increase within $\sim$ 3 au, and are much flatter beyond that.  This behavior change at $\sim$ 3 au could be indicative of the water-ice sublimation turn-on at this distance. A rapid increase in sublimation-driven activity is due to both ejection of dust particles from the surface and dragging of dust particles from the subsurface (through porous flow or disruptive mass wasting). Typically, our $Af\rho$ values are 2-3 times higher than those derived with spectroscopic or photometric instrumentation. This offset is most likely due to the relatively wide bandpass that visual observations use, which admit gaseous emission into the observer's field of view, such as from C$_2$, NH$_2$ and CO$^+$, especially closer to perihelion. 

As a reminder, the magnitudes used for this work were all obtained from visual observations by eye, sometimes with binoculars or telescopes, but always without filters or detectors. These $Af\rho$ values are likely to be contaminated by gas emission, which will lead to an overestimate of the comet's dust coma brightness, and thus,  higher $Af\rho$ values. Sky brightness, or less than ideal observing conditions would also lead to an underestimation of the coma diameter in a visual observation, which might contribute to the higher $Af\rho$ resulting from visual observations. Also, CCD magnitudes tend to be fainter than visual magnitudes, because they are obtained with short exposures to provide a good centroid for astrometry and avoid saturating the detector. A short exposure often means that the part of the coma will not be present in the photometric aperture thus leading to a lower magnitude. CCD observers also use telescopes with long focal length (vs. a visual observer's need for a short focal length) to achieve a low power and thus make the coma as small as possible to account for the low surface brightness that characterizes the outer part of comae. In contrast, the human eye is more likely to take in the whole coma + nucleus and hence the visual estimates tend to be brighter, leading to higher $Af\rho$ values. With future observations, this can be addressed by using narrow-band filter-imaging or spectroscopy where the dust contribution can be more accurately measured and calibrated (e.g.,  \citet{Milani2013}). 

\subsection{Correlation of Water and CO production rates with visual magnitudes}

We analyzed published CO and H$_2$O production rates \citep{biver02}, $Q({\rm CO})$ and Q(H$_2$O), respectively, in the context of the fifth-order polynomial fit to our $m_{\rm shift}$ magnitudes (Fig. \ref{fig:lcfits}). We wanted to test whether there was evidence for a possible effect of gas production on the visual magnitudes. In addition, this serves as an independent proof of concept that our approach can be readily combined with other data set to promote advanced analysis.

\begin{figure}
\centering
\includegraphics[width=.99\textwidth, height=0.5\textheight]{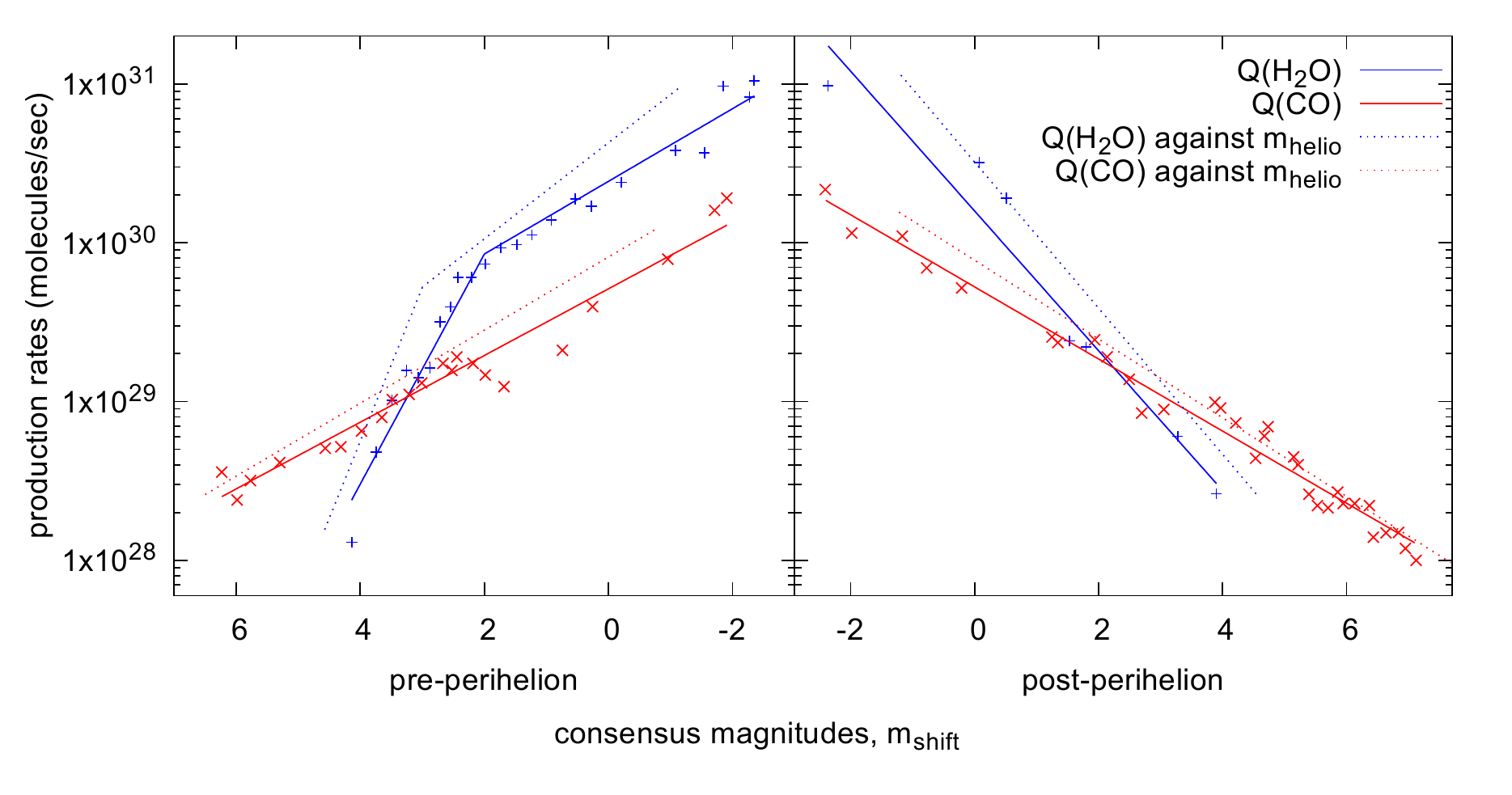}
\caption{Water and CO production rates in Hale-Bopp plotted against fifth order fit to final shifted magnitude (data and solid). Blue + signs are for Q(H$_2$O) and red x symbols are Q({\rm CO}). Pre- and post-perihelion data are plotted separately.  Dotted lines show fits if $m_{\rm helio}$ is substituted for $m_{\rm shift}$ in mapping distance to magnitude.
}
\label{fig:Correlations}
\end{figure}

As Fig. \ref{fig:Correlations} shows, the pre-perihelion water correlation slope has a noticeable break at $m_{\rm shift} \sim$ 2. Hale-Bopp was this bright predominantly when it was at $r \sim$ 2.6 au (see Figs. \ref{fig:mainlightcurve}d, \ref{fig:lcfits} and \ref{fig:final_logscale}). 

We derive these correlation equations for the two pre-perihelion H$_2$O regimes:

\medskip
\noindent $m_{\rm shift} \ge$ 2 (beyond $r \sim $ 2.6 au  pre-perihelion):
\begin{equation}
    \label{eq:H2Oouter}
\log_{10}(Q{\rm H_2O}) = 31.4-0.72m_{\rm shift}
\end{equation}

\noindent $m_{\rm shift} \le$ 2 (within $r \sim $ 2.6 au pre-perihelion):
\begin{equation}
\label{eq:H2Oinner}
\log_{10}(Q{\rm H_2O}) = 30.4 - 0.23m_{\rm shift}
\end{equation}

A similar change in the water-magnitude correlation was reported by \citet{BockeleeMorvanRickman}; however, they saw the break at $\sim$ 3 magnitudes, corresponding to $r \sim$ 3.1 au. We attribute the difference to the fact that they used $m_{\rm helio}$ values for their analysis. This point is illustrated in Fig. \ref{fig:Correlations} where we also show the correlation lines computed with our $m_{\rm helio}$ values with dotted lines, which reproduces \citet{BockeleeMorvanRickman}'s results. This change in correlation relationship for water production rate and magnitude may be useful in constraining models of Hale-Bopp activity; however, to get the most accurate identification of the heliocentric distance when this change occurred, we recommend applying a phase angle correction to the magnitudes. 

With the post-perihelion data, we find a single equation for H$_2$O:

\begin{equation}
\label{eq:H2Ocorrelationpost}
\log_{10}(Q{\rm H_2O}) = 30.2 - 0.44m_{\rm shift}, 
\end{equation}

\noindent which is similar to correlation relationships derived for other long-period comets \citep{sosafernandez2011,jorda2008}. 

In contrast, we find no break in the CO pre-perihelion correlation plot and derive a single equation:

\begin{equation}
\label{eq:COcorrelationpre}
\log_{10}(Q{\rm CO}) = 29.7 - 0.21m_{\rm shift}
\end{equation}

\noindent and a nearly identical relationship for post-perihelion:

\begin{equation}
\label{eq:COcorrelationpost}
\log_{10}(Q{\rm CO}) = 29.7 - 0.23m_{\rm shift}
\end{equation}

The pre- and post-perihelion correlation equations for CO (Eq. \ref{eq:COcorrelationpre}, \ref{eq:COcorrelationpost}) are remarkably similar, so that we recommend one equation for CO at all heliocentric ranges in comet Hale-Bopp: 

\begin{equation}
\label{eq:COcorrelationaveragemshift}
\log_{10}(Q{\rm CO}) = 29.7 - 0.22m_{\rm shift}
\end{equation}

The correlation data are consistent with the visual magnitudes increasing almost entirely due to CO outgassing until about 2.6 au, when the water also significantly contributes to the dust release. This coincides with the approximate heliocentric distance where water-ice sublimation began to dominate Hale-Bopp's activity.

\subsection{Estimating CO and H$_2$O Production Rates for Other Comets}

We can now take the correlation equations we derived for Hale-Bopp and apply it to other comets. If we note our assumption that the variance of the overall response of different comets is smaller than our uncertainties, we can proceed with caution. We may now use visual magnitudes of other new comets to predict what could be their associated gas production rates. Following this approach may be useful for planning observations of H$_2$O, CO and even other volatile species (OH, CN, HCN, CH$_4$, H$_2$CO, and CH$_3$OH), if we assume a range of typical cometary ratios for these species relative to water or CO.

Hence, we present another correlation equation for CO and $m_{\rm app}$, which we derived from our  $Q({\rm CO})$ and $m_{helio}$ correlation analysis for both post-perihelion and pre-perihelion Hale-Bopp data, and is plotted in red dotted line in Fig. \ref{fig:Correlations}:

\begin{align} 
\label{eq:COcorrelationaveragemhelio}
\begin{split}
\log_{10}(Q{\rm CO})  ~=&~ 29.9 - 0.24 m_{\rm helio}\\
~=&~  29.9 - 0.24(m_{\rm app} -5\log_{10}\Delta)\\
\end{split}
\end{align}

\noindent Although not corrected for phase angle\footnote{We do not correct for phase angle here, because accurate information about phase correction is unlikely to be known soon after a comet is discovered}, it should still be useful for a quick approximation of the CO production rate using using visual magnitudes, $m_{\rm app}$ from the ICQ, COBS, or Yoshida, and the comet's geocentric distance, $\Delta$. Of course, most comets will not be as productive as Hale-Bopp, but it is still often used as a comparison standard, especially for newly discovered or brighter than average comets.

As a demonstration, we consider a comet that has generated a great deal of interest recently: C/2017 K2 (PANSTARRS). Upon discovery it had a coma at very large heliocentric distances ($>$ 13 au), and its activity was hypothesized to be driven by CO outgassing \citep{meech17,jewitt17}. This Oort Cloud comet continues to brighten and is heading for a 2022 December perihelion. If we want to estimate its CO production rate for 2020 Jul 1, we use Eq. \ref{eq:COcorrelationaveragemhelio} and the reported apparent magnitude from that date, $m_{\rm app}$ = 16.4 (Minor Planet Electronic Circular 2020-M114) and obtaining $\Delta$ = 8.6 au (JPL Horizons), and predict that $Q({\rm CO})$ = 1.2x10$^{27}$ mol s$^{-1}$. A 3-$\sigma$ detection limit for this value would probably require more time than is typically granted at infrared and mm/submm-wavelength telescopes.

If C/2017 K2 (PANSTARRS)'s CO production rate matches that of Hale-Bopp, then we predict that it will be detectable at infrared and mm/submm wavelengths with  Q({\rm CO}) $\sim$ 10$^{28}$ mol s$^{-1}$ \citep{sen94,gun03,pag13,wierzchos20}, when it reaches $m_{\rm helio}  \sim$ 7.9. Based on recent lightcurve data from COBS or Yoshida, this is likely to occur during the first half of 2021 as the comet reaches $m_{\rm app} <$  12.5 and $\Delta$ = 8.5 au. It may be detected even sooner with more observing time. Of course, CO emission may still not be detectable in this comet, but such non-detections would also be valuable for constraining models of nucleus composition and activity. Thus, Eq. \ref{eq:COcorrelationaveragemhelio} may be useful for developing time estimates applicable in observing proposals for distantly active  comets.

Similarly, we present this equation from our correlation analysis of the post-perihelion water production rate and visual magnitudes for planning purposes:

\begin{equation}
\label{eq:H2Ocorrelationaverage}
\log_{10}(Q{\rm H_2O}) = 30.5 - 0.46(m_{\rm app} -5\log_{10}\Delta).
\end{equation}

\noindent This equation is plotted as a dotted blue line (post-perihelion) in Fig. \ref{fig:Correlations}. 

\bigskip

\section{Summary and Conclusions}

We present a detailed analysis of visual magnitudes for comet Hale-Bopp submitted to the International Comet Quarterly from more than a dozen observers, resulting in a lightcurve for each of these steps: 1) apparent magnitudes, 2) magnitudes corrected for geocentric distance only, 3) magnitudes corrected for both geocentric distance and phase angle, and 4) magnitudes corrected for geocentric distance, phase angle, and systematic observer variation.

Correcting Hale-Bopp's apparent magnitudes for geocentric distance brightens the lightcurve and changes its shape, especially around 3 - 4 au pre-perihelion. Correcting for light scattering, as measured by the comet's phase angle, changes the lightcurve slope by an average of $\sim$ 30\%. Not correcting for phase angle led to a reported flattening in the lightcurve from 4 to 2.6 au pre-perihelion, which was reported by other groups and attributed to the onset of water-ice sublimation near 3 au. However, we show that most of this reported flattening in this region was an artifact of not correcting for phase angle.

The self-consistent procedure resulted in adjustments to the slope roughly comparable to the original statistical error bars, while cutting the error bars approximately in half. The computer code used to derive the lightcurves, ICQSPLITTER, is available to the community for download on GitHub, for use on other comets. 

To first order, the pre-perihelion and post-perihelion lightcurves follow a $\sim$ $r^{-4}$ trend, which is consistent with typical trends often assumed for a relatively new comet. 

A closer look at the pre-perihelion data shows that the lightcurve deviates from a linear fit at these locations: 4.0, 2.6, 2.1 and 1.1 au. The 4.0 au inflection point coincided with the reported turn-on of four or more jets, or active areas, on the comet nucleus. The second change in slope at 2.6 au may coincide with the onset of vigorous sublimation of water ice from the nucleus. We are not aware of any other significant developments in Hale-Bopp's coma for the other changes in slope. 

The post-perihelion lightcurve was more linear than the pre-perihelion data, and was remarkably constant out to $\sim$ 9 au. 

The comet appeared to undergo an outburst a few days before perihelion, and two outbursts at $r$ = 2.15 and 7.36 au, post-perihelion.

The final absolute magnitudes derived from the lightcurve are $m_o$ = -1.74 and -1.92, which confirms that Hale-Bopp was one of the brightest comets in recent history.

The $Af\rho$ values derived from visual magnitudes in the final lightcurve show a similar power-law dependence on heliocentric distance, $r^{-1.5}$, as those derived using spectroscopic and imaging techniques, but the visually-derived $Af\rho$ values are 2-4 times larger.

We calculated and analyzed correlation laws between the published gas production rates (for H$_2$O and CO) and final shifted magnitudes.  The data are consistent with the visual magnitudes increasing almost entirely due to CO outgassing until $\sim$ 2.6 au, when the water begins to significantly contribute to the dust release. This coincides with the approximate heliocentric distance where water-ice sublimation began to dominate Hale-Bopp's activity.  

We present two equations to use for predicting water and CO production rates in comets using their apparent magnitude reports. Namely, 
\begin{align*}
\log_{10}(Q{\rm H_2O}) = 30.5 - 0.46(m_{\rm app} -5\log_{10}\Delta)
\end{align*}
\begin{align*}
\log_{10}(Q{\rm CO}) ~=&~  29.9 - 0.24(m_{\rm app} -5\log_{10}\Delta)
\end{align*}
\noindent

These relations have been validated for Hale-Bopp, which is a comet with prolonged and sustained activity that has been observed over a large heliocentric range. Further validation and improvement on the accuracy of such magnitude-production rate correlations will have to wait for future opportunities with similarly active objects. We suggest these relations be used as a quick reference for guiding modeling work and proposal planning.

\bigskip
\bigskip

This material is part of the  ``Cometwatchers Lightcurve Project,'' carried out with undergraduate researchers at St. Cloud State University and the University of South Florida, supported in part by the National Science Foundation under Grants No. AST-1615917, and AST-1945950. This research has made use of data and/or services provided by the International Astronomical Union's Minor Planet Center. M.W.\ acknowledges very helpful discussions with  D.W.E. Green.  This material is based in part on work done by D.A.R.\ while serving at the National Science Foundation.

\bibliography{bibtex.bib}



\appendix
\section{ICQSplitter} 
\label{sec:ICQSplitter}

The International Comet Quarterly Splitter (ICQSplitter)$^3$ is a Python based open-source software which will take data from the ICQ$^1$,

Comet OBServation Database$^5$(COBS), and JPL HORIZONS\footnote{\url{https://ssd.jpl.nasa.gov/horizons.cgi}} to produce lightcurves of a specified target. The pipeline can be run on Unix or Windows-based operating systems. ICQSplitter was used in this text to produce lightcurves of visual magnitude data from amateur astronomers, but it is capable of taking in any measurements, including those from charge-coupled devices (CCD) that are reported in ICQ's standard 80-column format. The user has the options to apply any combination of corrections discussed in the main body of this paper. For example, a user with observational magnitudes from a relatively non-dusty comet may wish to forgo the application of a phase correction.

ICQSplitter is available for download on GitHub.
Since ICQSplitter is still evolving, refer to the on-line documentation for up-to-date information. This document describes the functionality of ICQSplitter Version 3.0 as of 28 January 2020. Also refer to the documentation for installation guides and additional support.

\section*{1.1 Methods and Implementation}

This software is a standalone Python\footnote{\url{https://www.python.org/}}  3.6.4 script \citep{van2011python}. It makes use of Python packages that are freely available and easy to install through the Python Package Index;\footnote{\url{https://pypi.python.org/pypi}} required packages include NumPy \citep{van2011numpy}\footnote{\url{http://www.numpy.org/}}, SciPy \citep{jones2014others}\footnote{\url{https://www.scipy.org/}}, matplotlib \citep{hunter2007matplotlib}\footnote{\url{http://matplotlib.org/}}, and CALLHORIZONS \citep{mommert2017photometrypipeline}\footnote{\url{https://github.com/mommermi/callhorizons}}. At its base level (i.e., without any optional command line arguments input) this program will read in ICQ or COBS 80 column format and filter out data that do not meet the criteria discussed in \ref{sec:removed}. The data must be from a a single small body comprised of either entirely CCD or visual magnitude data and comprised of observations from a single orbit around the sun spanning a date range no larger than five years.

\section*{1.2 ICQSplitter Arguments}

ICQSplitter has a few optional command line arguments. Refer to the documentation provided online for more details on how to use them.

\section*{1.2.1 --heliocentric} \label{JPL}

This command applies heliocentric corrections to the raw magnitudes. In doing so, ICQSplitter will use the CALLHORIZONS package to query JPL HORIZONS to extract the heliocentric distance, geocentric distance, and phase angle of the target. This function will perform a single query of JPL HORIZONS over the range of dates provided in increments inputted by the user. The default time interval is 30 minutes increments. For instance,  if your first date is 1996:01:19 00:00, final date is 1996:01:19 01:00, and your increment size is every 30 minutes then it will query JPL Horizons for the ephemerides of your object at 1996:01:19 00:00, 1996:01:19 00:30, and 1996:01:19 01:00. As JPL only allows users to pull 100,000 in a single query, if one wishes to analyze all of their data in a single compilation of ICQSplitter then small time increments should be avoided. ICQSplitter uses the ephermides data to perform heliocentric corrections following equations discussed in section \ref{sec:params}. 

\section*{1.2.2 --phase} \label{schleicher}

Applies phase corrections to the given magnitudes. If --heliocentric and --phase are called at the same time then the phase angle corrections will be applied onto the heliocentric corrected magnitudes as discussed in section \ref{sec:params}, else JPL will be queried for the first time and phase corrections will be applied to the raw magnitudes. The phase angles are cross referenced to Dave Schleicher's Composite Dust Function for Comets \citep{schleicher11}.

\section*{1.2.3 --stats} \label{stats}

Performs the statistical analysis discussed in section \ref{sec:stats}. The program will automatically split any dataset into pre- and post-perihelion and perform the statistics on each set separately. ICQSplitter follows procedures for regression analysis through the methods of singular value decomposition using NumPy's Linear Algebra package. After a polynomial fit has been taken to convergence, Python's Statistics package is used to perform the Students t and probability tests on each observer's data. If one observer is found to fail the stationarity test in either epoch, then that observer is removed from the dataset and the procedure is repeated. The --stats command is always issued after --heliocentric and --phase (if those commands have also been given). 

\section*{1.2.4 --plot} \label{plots}

This command will produce plots with Pythons matplotlib package. For instance, if a user runs her code on data with the --phase and --stats commands then --plots will produce individual graphs of $m_{\rm app}$, $m_{\rm helio}$, $m_{\rm phase}$, and $m_{\rm shift}$.

\end{document}